\documentclass[referee]{aa}

\usepackage{natbib,twoopt}
\usepackage{graphicx}
\usepackage{float}
\usepackage{caption,subcaption}
\usepackage{txfonts}
\usepackage{multirow}
\usepackage{tablefootnote}
\usepackage{etoolbox}

\makeatletter
\patchcmd\@combinedblfloats{\box\@outputbox}{\unvbox\@outputbox}{}

\makeatother

\newcommand{\comment}[1]{}

\begin{document} 
   \title{Machine learning technique for morphological classification of galaxies from SDSS}
   \subtitle{II. The image-based morphological catalogues of galaxies \\at 0.02<z<0.1}

  \author{Vavilova I.B.$^{1,2}$, Khramtsov V.$^{3}$,  Dobrycheva D.V.$^{1}$, Vasylenko M.Yu.$^{1,4}$,  \\Elyiv A.A.$^{1}$, Melnyk O.V.$^{1}$}

   \institute{
$^{1}$Main Astronomical Observatory of the NAS of Ukraine, 27, Akademik Zabolotny Str., Kyiv, 03143, Ukraine\\
$^{2}$Astronomical Observatory of the I.I. Mechnikov National University of Odesa, 1v, Marazliyivska Str., Odesa, 65014, Ukraine\\
$^{3}$Institute of Astronomy, V.N. Karazin Kharkiv National University, 35 Sumska Str., Kharkiv, 61022, Ukraine\\
$^{2}$Institute of Physics of the NAS of Ukraine, Nauka av., 46, Kyiv, 02000, Ukraine}

\titlerunning{Image-based morphological catalogues}
\authorrunning{Vavilova I.B., Khramtsov V., Dobrycheva D.V. et al.}

   \date{Received September 21, 2020; Revision Received August 1, 2021}
   
   \abstract
{We applied the image-based approach with a Convolutional Neural Network (CNN) model to the sample of low-redshift galaxies with $-24^{m}<M_{r}<-19.4^{m}$ from the SDSS DR9. We divided it into two subsamples, SDSS DR9 galaxy dataset and Galaxy Zoo 2 (GZ2) dataset, considering them as the inference and training datasets, respectively. To determine the principal parameters of galaxy morphology defined within the GZ2 project, we classified the galaxies into five visual types (completely rounded, rounded in-between, smooth cigar-shaped, edge-on, and spiral). Using GZ2 galaxy morphology classification, we were able to define 34 morphological features of galaxies from the inference set of our SDSS DR9 sample, which do not match with the GZ2 training set. As a result, we created the morphological catalogue of 315\,782 galaxies at $0.02<z<0.1$, where morphological five classes were first defined for 216\,148 galaxies by image-based CNN classifier. For the rest of galaxies, the initial morphological classification was re-assigned as in the GZ2 project. 

Our method shows the promising performance of morphological classification attaining >93\,\% of accuracy for five classes morphology prediction except the cigar-shaped ($\sim75\,\%$) and completely rounded ($\sim83\,\%$) galaxies. Main results are presented in the catalog of 19468 completely rounded, 27\,321 rounded in-between, 3\,235 cigar-shaped, 4\,099 edge-on, 18\,615 spiral, and 72\,738 general low-redshift galaxies of the studied SDSS sample.
   
As for the classification of galaxies by their detailed structural morphological features, our CNN model gives the accuracy in range 92–99\,\% in depending on features, number of galaxies with the given feature in the inference dataset, and, of course, the galaxy image quality. As a result, for the first time we assigned 34 morphological detailed features (bar, rings, number of spiral arms, mergers, etc.) for more than 160\,000 low-redshift galaxies from the SDSS DR9. 
   
We demonstrate that implication of the CNN model with adversarial validation and adversarial image data augmentation improves classification of smaller and fainter SDSS galaxies with $m_{r}$ <17.7. The proposed CNN model allows solving a bunch of galaxy classification problems, for example, such as a quick selection of galaxies with a bar, bulge, ring, and other morphological features for their subsequent analysis.
}
   \maketitle

\section{Introduction}

Since the beginning of extragalactic astronomy and the first catalogues of galaxies by Ch. Messier, F.W. Hershel and J.F.W. Hershel, J.L.E. Dreyer, the image-based morphological classifications of galaxies have played a vital role in reflecting the evolutionary history of various types of galaxies and the large-scale structure of the Universe as a whole (\cite{Davis1985, Peebles1993, Barrow1993, Yang2003, Bundy2010, Peng2010, Reid2012, Leung2019}.

Manual galaxy morphological classification as the most precise method requires extensive usage of human resources, either from highly skilled professionals or, in some cases, amateur astronomers and volunteers such as in the Galaxy Zoo (GZ) project \cite{Willett2013}. Current and near-term galaxy observational surveys as the SDSS, LSST, DES, KiDS, SKA, the Euclid satellite, JWST, etc., are approaching the Exabyte scale multiwavelength databases of hundreds of millions of galaxies, which is impossible to classify manually. For instance, Vera C. Rubin Observatory (LSST), which will be operated starting from 2022, is expected to generate about 30 TB of data per night, revealing $\sim20$ million galaxies over this time \cite{Ivezi2019}, more than the Sloan Digital Sky Survey (SDSS) over its lifetime \cite{Blanton2017}. It is also worth mentioning the Euclid survey, which aims to detect billions of galaxies over 15\,000 square degrees of the celestial sky \cite{Amiaux2012}, and other big data astronomical projects (see, for example, a recent review of multiwavelength surveys and catalogues by Vavilova et al. \cite{Vavilova2020c}). Moreover, the human mind is not able to comprehend complex correlations in the diverse space of parameters, and multidimensional mathematical analysis is the best tool for determining the various common features between different types of objects. All that exaggerates the interest to use the alternatives in the form of machine learning (ML) techniques, including deep learning (DL), for the classification, forecasting, and discovery of various properties of galaxies (see, for example, \cite{Brugere2016, Baron2019, Fluke2020, Vavilova2020a}.

In this context, we note several recent prospective applications of Convolutional Neural Networks (CNNs) to classify the galaxies by their different parameters.

\cite{Cabayol2021} have demonstrated the CNN capability to avoid distorting effects when extracting the galaxy photometry from astronomical images as Lumos architecture. Exploiting the PAU imaging survey, these authors combined a CNN and a Mixture Density Network that allowed them to measure the photometry of a blended galaxy with the high accuracy. \cite{Diego2020}, in their work with DL in classifying early- and late-type galaxies in the OTELO and COSMOS databases, have used optical and infrared photometry and available shape parameters (the Sérsic index or the concentration index). Regardless of slight differences in the photometric bands used in each catalog, their neural network architecture operates well with missing data. 

The distance moduli and photometric redshift estimates benefit from the ML utilization into the big data sets, which provide a wide number of galaxy features for learning. \cite{Pasquet2018} used DL for classifying, detecting, and predicting photometric redshifts of quasars in SDSS. In works by \cite{Kugler2016, Speagle2017, Disanto2018, Salvato2019, Elyiv2020}, the machine learning methods were applied to assign and predict photometric redshifts within large-scale galaxy surveys with good accuracy. The GAN approach serves as a basis for restoring galaxy distribution in the Zone of Avoidance (\cite{Schawinski2017, Vavilova2018}, and generating dark matter structures in cosmological simulations (\cite{Diakogiannis2019}. 

Among the CNNs modeling in tasks of multiwavelength sky surveys, we note as follows: search for blazar candidates in the Fermi-LAT Clean Sample \cite{Kang2019}; boosted decision tree for detecting the faint gamma-ray sources with future Cherenkov Telescope Array \cite{Krause2017, Ruhe2020}; infrared color selection of Wolf-Rayet star candidates in our Galaxy using the Spitzer GLIMPSE catalogue \cite{Morello2018}; cosmic string searches in 21-cm temperature CMB maps \cite{Ciuka2017}; neural network-based Faranoff-Riley classifications of radio galaxies from the Very Large Array archive \cite{Aniyan2017}; deep learning classification of compact and extended radio source from Radio Galaxy Zoo project \cite{Lukic2018}; CNN for morphological assignment to radio-detected galaxies with active nuclei \cite{Ma2019}. \cite{Scaife2021}, in recent work, presented the first application of group-equivariant CNN to radio galaxy and their image translations, rotations, and reflections \cite{Scaife2021}.

Deep learning is promising for generating various synthetic catalogues and mock images, which helps to interpret the observational data \cite{Khalifa2017} and to discover new galaxies as, for example, high-z «blue nuggets» from the CANDELS survey \cite{Huertas2018}; to reveal structural properties of dark matter halos to their assembly history and environment \cite{ChenYangyao2020}; to establish a topology of the large-scale structure of the Universe in DM cosmological simulation \cite{Tsizh2020}; to separate the radiation from active galactic nuclei and star-forming galaxies with recognition method based on Deep Neural Network \cite{Chen2021}.

As for the discovery of new classes of celestial bodies with CNN, we highlight the works related to the gravitational lenses \cite{Jacobs2019, Khramtsov2019b} and the transient events and objects as supernovae, gamma-ray bursts, jets, etc. For example, the Catalina Real-Time Transient Survey serves as the platform for their detection and monitoring (see, for example, \cite{Mahabal2011, Djorgovski2016,  Mahabal2019}) as well as the Zwicky Transient Factory \cite{Bellm2019} as the LSST precursor and ML models implementation in synoptic sky surveys.

The CNN models have played a crucial role in analyzing data streams from the Advanced Laser Interferometer Gravitational-Wave Observatory (LIGO) detectors allowing to register gravitational wave signals from coalescing black hole binaries. Among such works are ones on the training an ML system for real-time glitch classification \cite{Zevin2017}, on the classification of gravitational wave signals, events, and instrumental noise \cite{George2018a, George2018b} as well as to test the theories on binary black hole mergers upon which the models are based \cite{Huerta2018}.

So, CNNs reliably manage with tasks for a variety of image-based classification, regression, prediction, and discovery of galaxies and other celestial bodies. 

In our work, we used a deep CNN model for the image-based morphological classification of $\sim300\,000$ galaxies ($0.02<z<0.1$) from SDSS DR9. To do this, we divided galaxies by their images \cite{Zhu2019} into five morphological classes (completely rounded, rounded in-between, cigar-shaped, edge-on, and spiral galaxies) as in the Galaxy Zoo 2 (GZ2) project. In our previous works \cite{Khramtsov2019a, Vasylenko2020}, we used binary classification but, as it turned out, it is difficult to correctly divide galaxies into two classes using the assigned label of galaxies from the GZ2. We investigate the problem of differences in these datasets and suggest ways to overcome adversarial validation. We also used our CNN model to predict 34 detailed structural morphological features (bar, ring, bulge, mergers, number of spiral arms etc.) of these galaxies with ware labeled in the GZ2 project \cite{Walmsley2020}. 

The structure of our paper is as follows. The sample of galaxies is described in Section 2. CNN model as the image morphological classifier, training and inference galaxy datasets are presented in Section 3. The created galaxy catalogues and results are given in Section 4 as well as discussion and conclusion in Sections 5 and 6, respectively. 

\section{Sample of low-redshift galaxies from the SDSS}

We used a representative sample of the 316031 SDSS galaxies at $0.02<z<0.1$ (with velocities correction on the velocity of Local Group, $V_{LG}$>1500 km/s). This sample was studied by us practically as "galaxy by galaxy" in previous works (\cite{Chesnok2009, Dobrycheva2013, Dobrycheva2014, Dobrycheva2015, Dobrycheva2017, Dobrycheva2017a, Elyiv2009, Vol2011, Melnyk2012, Pulatova2015, Vasylenko2019, Vasylenko2020, Vavilova2005, Vavilova2009, Vavilova2020a, Vavilova2021e}). Our most previous  research was to apply the ML photometry-based approach for binary morphological classification of these galaxies \cite{Vavilova2020a} and to create the catalog of their morphological types (early and late) obtained with the Support Vector Machine with an accuracy of 96.4 \% (\cite{Vavilova2021b}).

The main stages in preparing this sample were as follows. A preliminary set of galaxies at $z<0.1$ with the absolute stellar magnitudes $-24^{m}<M_{r}<-13^{m}$ from the SDSS DR9 contained $\sim 724\,000$ galaxies. Following the SDSS recommendation, we limited the sample to $m_{r}$ < 17.7 in \textit{r}-band to avoid typical statistical errors in spectroscopic flux. After excluding the images with stars and artifact objects as well as the duplicates of galaxy images, the final sample consisted of \textit{N} = 315\,782 galaxies. To clear the sample from segmented images of the same galaxy, we used our code based on the minimal angle distances between such SDSS objects. 

\begin{figure}
	\includegraphics[width=1.0\textwidth]{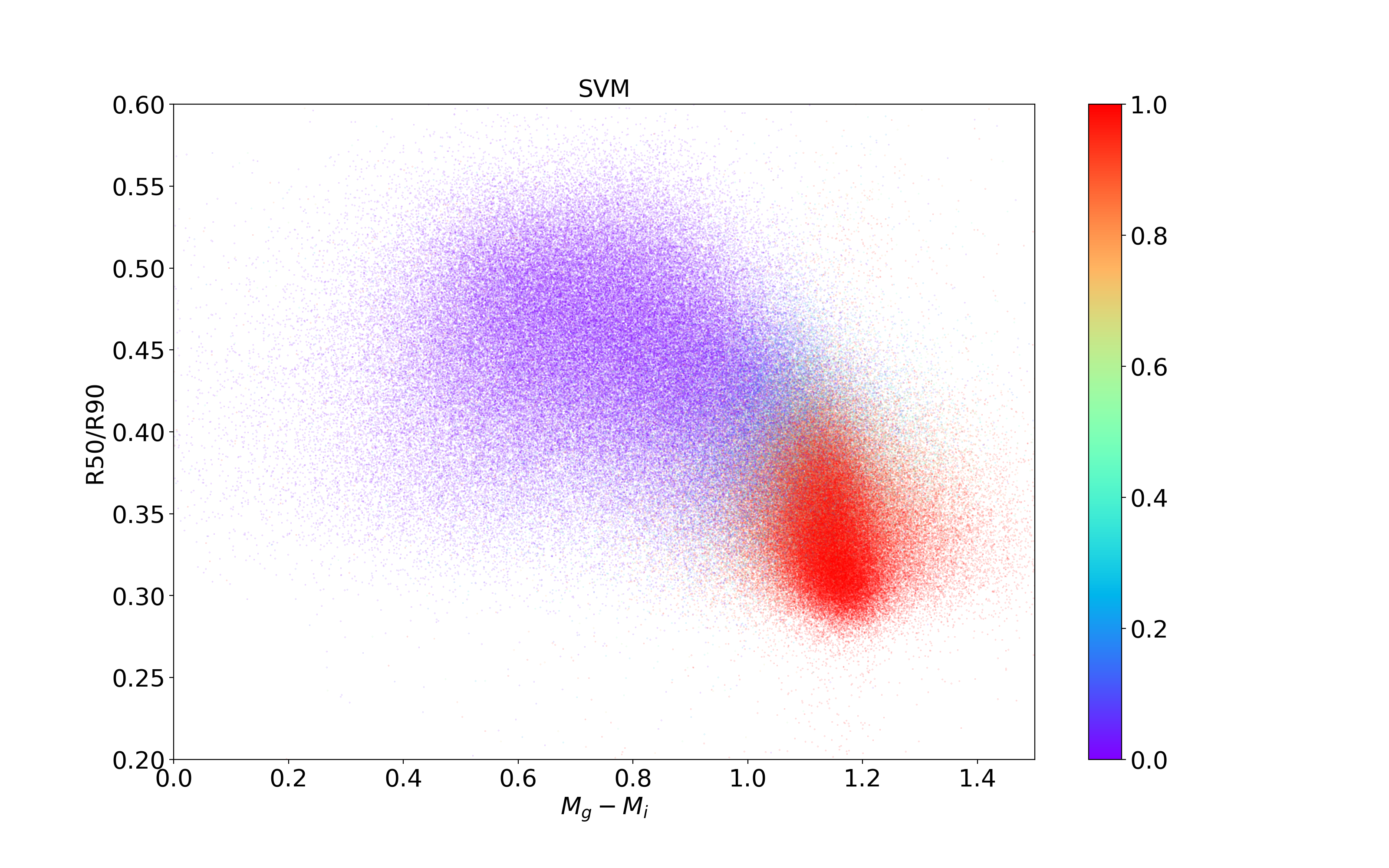}
    \caption{Diagram of color indices $g-i$ and inverse concentration indexes $R50/R90$ of the studied low-redshift galaxies from the SDSS DR9 after applying the Support Vector Machine (SVM) method: red color –- early $E$ (from elliptical to lenticular) and blue color –- late $L$  (from S0a to irregular Im/BCG) morphological types. Color bar from 0 to 1 shows SVM probability to classify galaxy as the late to the early morphological type [98].}
    \label{SVM}
\end{figure}

The absolute stellar magnitude of the galaxy was obtained by the formula 
$$
M_{r}=m_{r}-5\cdot\lg(D_{L})-25-K_{r}(z)-ext_{r},
$$
where $m_{r}$ is the stellar magnitude in \textit{r}-band, $D_{L}$ is the luminosity distance, 
$ext_{r}$ is the the Galactic absorption in \textit{r}-band in accordance to \cite{Schlegel1998},  
$K_{r}(z)$ is the cosmological k-correction in \textit{r}-band according to \cite{Chilingarian2010, Chilingarian2012}. The color indices were calculated as 
$$M_{g}-M_{i} = (m_{g}-m_{i}) - (ext_{g} - ext_{i}) - (K_{g}(z) - K_{i}(z)),$$
where $m_{g}$ and $m_{i}$ are the stellar magnitude, $ext_{g}$ and $ext_{i}$ are the Galactic absorption, $K_{g}(z)$ and $K_{i}(z)$ are the cosmological k-corrections in \textit{g} and \textit{i} bands, respectively. 

We found that Support Vector Machine gives the highest accuracy exploiting different galaxy classification techniques: human labeling, multi-photometry diagrams, and five supervised ML methods. Namely, it attains 96.1 \% for early E and 96.9 \% for late L types of galaxies \cite{Vavilova2020a}. We verified dependencies between accuracy and redshifts, human labeling bias, the overlap of different morphological types for galaxies with the same color indices, edge-on and face-on galaxy shape to determine the ability of each method to predict the galaxy morphological type. Distribution of 315782 galaxies from the studied SDSS sample by their morphological type (early and late) is given in Fig.~\ref{SVM}.

\section{CNN models for image-based morphological multi-label classification of galaxies}

\begin{figure}
	\includegraphics[width=1.0\textwidth]{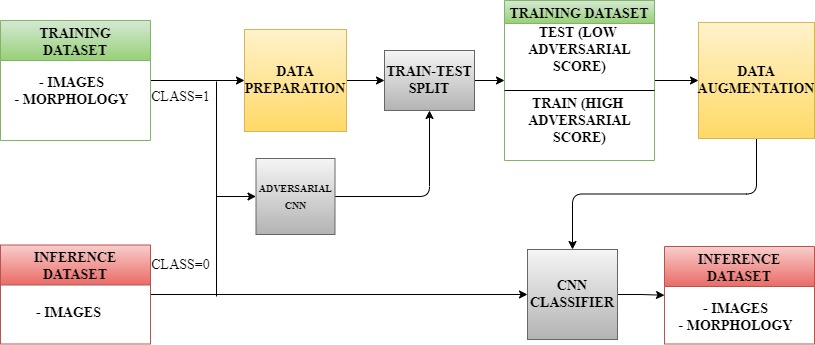}
    \caption{Block-diagram of the image-based classification of galaxies with CNN model for five morphological classes. The algorithm consists of the training/inference datasets reduction, image data processing, adversarial validation, special train-test split via adversarial scores, data augmentation, CNN classifier.}
    \label{chart}
\end{figure}

The studied sample of 315\,782 galaxies from SDSS DR9 is tightly overlapped with the data from the Galaxy Zoo 2, GZ2 \cite{Willett2013}. It allows us to divide it into two datasets: ``inference dataset'' of 143\,410 galaxies, which do not match  the GZ2 dataset; ``training dataset'' of 172\,372 galaxies, which match galaxies from our studied sample. The block-diagram of the image-based classification of galaxies with our CNN model is given in Fig.~\ref{chart}). 

For each galaxy from both datasets we have their SDSS images, but morphological classes are defined only for galaxies from the training GZ2 dataset. The 315782 RGB images of galaxies were requested from the SDSS (\url{http://skyserver.sdss.org/dr15/en/help/docs/api.aspx##cutout}). They are composed of $gri$-bands \cite{Lupton2004} having color scaling, each of 100$\times$100$\times$3 pixels ($39.6\times39.6$ arcsec) in each channel of the RGB image. 

\subsection{Galaxy morphological image classification into five classes by shape}

It is important to note the principal difference between the galaxy images in our inference dataset and the GZ2 training dataset. Galaxies from the inference dataset are much shallower than those from the GZ2 dataset. As we mentioned in Section 2, the galaxies from the studied sample were pre-selected via $m_{r}$ < 17.7 limit following the SDSS recommendation. This affects the value of the 90\,\% Petrosian flux parameter. Thereat, the galaxies, which do not match the GZ2 dataset, are fainter and smaller on average than galaxies from the training GZ2 dataset. To get around this problem, we used an adversarial CNN to compare these two datasets (training and inference). Namely, we trained it on all galaxy images of our sample, passing the class `0' for the inference dataset and class `1' for the training dataset (Fig. \ref{chart}).
To develop the CNN model based on the images of galaxies, we used the GZ2 assigned labels for five morphological classes by shape: completely rounded, rounded in-between, cigar-shaped, edge-on, and spiral galaxies. 

We provided additional data cleaning of 172\,372 galaxy images from the training sample and took into consideration only those galaxies for which GZ2's volunteers gave the most votes for a more accurate result (Fig. \ref{chart}). It turned out to be 72738 galaxies. The criteria for each image of the galaxy were defined in the GZ2 project \cite{Willett2013}; their description is available through the web-site \url{https://data.galaxyzoo.org/}. The criteria with the \textsc{*\_count} prefix indicate the number of votes of volunteers; other criteria correspond to the debiased fraction of votes assigned in the GZ2 catalogue with the \textsc{*\_debiased} prefix. So, we applied criteria for galaxies belonging to the five morphological classes by shape as follows: 
\begin{itemize}
    \item completely rounded: \textsc{smooth} (number of votes > 0.469), \textsc{completely\_round} (>0.469), \\ \textsc{smooth\_count} (>25), \textsc{completely\_round\_count} (>25);
    \item rounded in-between: \textsc{smooth} (>0.469),  \textsc{rounded\_in\_between} (>0.5), \textsc{smooth\_count} (>25),\\  \textsc{rounded\_in\_between\_count} (>25);
    \item cigar-shaped: \textsc{smooth} (>0.469), \textsc{cigar\_shaped} (>0.5), \textsc{smooth\_count} (>25),\\ \textsc{cigar\_shaped\_count} (>25);
    \item edge-on: \textsc{features\_or\_disk} (>0.43), \textsc{edgeon\_yes} (>0.602), \textsc{features\_or\_disk\_count} (>25), \\ \textsc{edgeon\_yes\_count} (>25);
    \item spiral: \textsc{features\_or\_disk} (>0.43), \textsc{edgeon\_no} (>0.715),  \textsc{spiral} (>0.619), \\ \textsc{features\_or\_disk\_count}  (>25), \textsc{edgeon\_no\_count} (>25), \textsc{spiral\_count} (>25). 
\end{itemize}

These criteria with vote scores >0.4 and higher were found to be quite good for providing a reliable image-based morphological classification of galaxies. The galaxy image data preparation and augmentation for our CNN model is described in detail in our paper \cite{Khramtsov2022}.

The adversarial CNN resulted in the fact that the training dataset contains galaxies, which properties are not common with the inference dataset. The useful parameter to solve this task is the adversarial score, which means the probability of the galaxy with some feature to be similar to the galaxy with the same feature from the GZ2 training dataset. So, we can test galaxies with a low adversarial score from the training dataset in a way to train them on galaxies with the high adversarial score from the training set (Fig.~\ref{chart}). On the step of this ``train-test split'', we randomly selected $\sim 9000$ galaxies with an adversarial score less than 0.7 to test the CNN classifier (test-split training dataset), and the rest part of this dataset ($\sim 63\,000$) to train the CNN classifier (train-split training dataset). Because this score is related to the presence of fainter and smaller galaxies in the inference galaxies, we have done the image data augmentation of galaxies from the training dataset. Namely, we decreased the intensity of pixels and reduced the image size. Distribution of predicted labels vs. true labels of five morphological classes for these $\sim 9\,000$ galaxies with the high adversarial score as the confusion matrix is shown in Table~\ref{classes}. One can see that our model for CNN classifier guarantees >93\,\% of accuracy for rounded in-between, edge-on, and spiral morphological classes, 83\,\% for the completely rounded and 75\,\% for the cigar-shaped galaxies.

\begin{table*}[]
    \centering
\begin{tabular}{l|c|c|c|c|r}
\hline
& Predicted labels \\ \hline
True labels	&Completely & Rounded & Cigar-shaped & Edge-on	& Spiral\\
 & rounded & in-between &  & & \\
\hline
Completely rounded & \textbf{0.83} &0.16 & 0 &	0.00038	& 0.012 \\
Rounded in-between &0.054 &\textbf{0.93} &0.0047 &	0.00025 &0.015 \\
Cigar-shaped &0	& 0.17 &\textbf{0.75} &0.065 &0.017\\
Edge-on& 0&	0.0076&	0.049&	\textbf{0.93} & 0.0092\\
Spiral&	0.0075&	0.022& 0.0017 &	0.0092 &\textbf{0.96}\\
\end{tabular}
\caption{Distribution of predicted labels vs. assigned labels as the probabilities for galaxy to belong to one of five morphological classes (CNN classifier for test-split training galaxy dataset of 9000 galaxies). Each row represents the fraction of galaxies from a certain class (defined at the horizontal row) to be classified as galaxies from other classes.}
    \label{classes}
\end{table*}

As a result of our CNN classifier, we got the morphological classes of 72\,738 galaxies from the training set as follows: 19\,468 completely rounded, 27\,321 rounded in-between, 3\,235 cigar-shaped, 4\,099 edge-on, and 18\,615 spiral galaxies.

Meanwhile, knowing the morphological class of galaxies from the training dataset, we are able to classify galaxies from the inference dataset with CNN. We compared a few CNN models for the five-class morphological classification. Following our previous works \cite{Khramtsov2019a, Vasylenko2020} the best neural network for our task is DenseNet-201. 

Our CNN model consists of two main parts. The first one is the convolutional part, where CNN performs the image processing with a gradually decreasing size. The highlight of CNN architectures is to use the fully connected layers at the tail. This tail corresponds to the neural network classifier, which transforms the output of the convolutional part into the dense layer, the number of neurons in which is equal to the number of classes (see, for example, \url{http://cs231n.stanford.edu/}). So, the second part of our model is the fully connected part, where the processed galaxy image comes through a few layers of connected neurons up to the last layer, consisting of five neurons. The output of the last layer corresponds to the probability of a galaxy being one of five defined classes.

Our CNN model was completed by the two dense layers of neurons (with the number of neurons equal 128 and 5, respectively) and, after, by the global max-pooling. The activation functions at the tail of the CNN model were the same as in adversarial validation. As an optimizer, we used the ADAM with an initial learning rate of 1$\times$10$^{–4}$; the optimizer minimized the \textit{categorical crossentropy} loss function (see, in detail, \cite{Khramtsov2022}). 

\section{Galaxy morphological multi-label classification by 34 features}

Besides classification into five morphological classes, the image galaxies from the training dataset attribute 37 parameters of the detailed morphology. They are estimated for each galaxy according to volunteers' answers and form the decision tree \cite{Richert2013, Willett2013}. The principal restriction for classification with CNN is the presence of the only parameter, which characterizes this class of objects on an image \cite{LeCun2006}. We introduced the ``similarity learning'' approach: if two galaxies have similar images, then their morphological parameters are similar. In other words, we exploited the galaxy images from the training dataset, which are most similar to the galaxy images from the inference dataset by their 37 morphological detailed features. The algorithm is finding the nearest galaxies from the training dataset to the target galaxy of the inference dataset in the penultimate CNN layer of multi-parameter space and assigns the attributes of the nearest neighboring galaxies to the target galaxy.

We also applied the adversarial validation to predict 37 detailed morphological features of galaxy images from the inference dataset with some adversarial score. The three very sparse features were removed from the consideration. So, we worked with the inference dataset of 160\,471 galaxies and with 34 galaxy morphological features (bar, ring, various number of spiral arms, disks, dust lane, merger, etc.). 

These morphological features are listed in Table~\ref{features}. The names of features (“parameter”) are given in the first column as they are labeled by the GZ2’s volunteers. The numbers of galaxies in the inference dataset with the given feature are given in the last column 5. As the binary classification quality metric we used \textit{Receiver Operating Characteristics}, which is determined with \textit{Area Under Curve} quantitate value (ROC AUC). The columns 2–3 contain the ROC AUC for galaxies of GZ2 dataset (ROCtest). Two resulting accuracy scores, measured with the ROC AUC classification quality metric, give the score for the model trained with adversarial augmentations (ROCtest AUG, column 3) and for the model, trained without these augmentations (ROCtest NOAUG, column 2). To estimate the ROC AUC, one needs to plot the following curve: the fraction of true positives out of the positives (TPR = true positive rate) versus the fraction of false positives out of the negatives (FPR = false positive rate) at various threshold settings to estimate the area under this curve. The ROC AUC equals 1 for ideal classification and 0.5 for random one. For more information about ROC AUC classification quality metric measuring see the paper by Bradley \cite{Bradley1997}. The column 4 gives the difference between ROCtest AUG and ROCtest NOAUG values.

One can compare these scores and estimate the degree of influence of image data augmentations on the classification quality of a trained model. Scores are given for the dataset of $\sim 9\,000$ galaxies, expanded with a fraction of galaxies, which do not pass the criteria of morphological classification. 

There is a particular discrepancy in the numbers of galaxies with detailed morphological features from the training dataset. It can be explained, among other things, by the fact that the GZ’ volunteers did not notice certain morphological details while the CNN classifier found.

\begin{table*}[]
    \centering
\begin{tabular}{l|c|c|c|r}
\hline
Feature’s name [109]
 & ROC$^{\text{test}}_{\text{NOAUG}}$ & ROC$^{\text{test}}_{\text{AUG}}$ & ROC$^{\text{test}}_{\text{diff}}$ & 
 Number of galaxies\\
 & & & & in inference dataset\\
               \hline    
\textsc{smooth} & 89.25\% & 88.59\% & -0.66\% & 624\\
\textsc{features\_or\_disk} & 92.54\% & 91.88\% & -0.66\% & 19\,770 \\
\textsc{star\_or\_artifact} & 95.36\% & 97.63\% & 2.28\% & 6 \\
\textsc{edgeon\_yes} & 98.81\% & 98.65\% & -0.16\% & 2\,079 \\
\textsc{edgeon\_no} & 97.21\% & 96.82\% & -0.39\% & 7\,504 \\
\textsc{bar} & 93.99\% & 92.41\% & -1.57\% & 90 \\
\textsc{no\_bar} & 90.69\% & 89.80\% & -0.90\% & 1\,762\\
\textsc{spiral} & 93.40\% & 92.88\% & -0.52\% & 1\,199 \\
\textsc{no\_spiral} & 86.30\% & 84.78\% & -1.52\% & 47 \\
\textsc{no\_bulge} & 98.36\% & 98.35\% & -0.01\% & 63 \\
\textsc{odd\_yes} & 94.78\% & 93.37\% & -1.41\% & 1\,096\\
\textsc{odd\_no} & 84.62\% & 83.51\% & -1.11\% & 61\,537 \\
\textsc{completely\_round} & 96.17\% & 95.60\% & -0.58\% & 6\,018 \\
\textsc{rounded\_in\_between} & 92.31\% & 91.46\% & -0.85\% & 20\,107 \\
\textsc{cigar\_shaped} & 97.96\% & 97.73\% & -0.23\% & 12\,434 \\
\textsc{ring} & 96.97\% & 96.43\% & -0.54\% & 52 \\
\textsc{irregular} & 96.74\% & 96.94\% & 0.20\% & 41 \\
\textsc{other} & 95.93\% & 89.20\% & -6.74\% & 4 \\
\textsc{merger} & 91.79\% & 88.89\% & -2.90\% & 8 \\
\textsc{dust\_lane} & 99.39\% & 99.40\% & 0.02\% & 4\\
\textsc{bulge\_shape\_rounded} & 96.73\% & 96.27\% & -0.47\% & 18 \\
\textsc{bulge\_shape\_no\_bulge} & 98.65\% & 98.52\% & -0.13\% & 664\\
\textsc{arms\_winding\_tight} & 89.45\% & 88.60\% & -0.85\% & 3\\
\textsc{arms\_winding\_medium} & 75.33\% & 77.59\% & 2.26\% & 2\\
\textsc{arms\_winding\_loose} & 94.95\% & 94.41\% & -0.54\% & 100\\
\textsc{arms\_number\_2} & 90.55\% & 89.99\% & -0.56\% & 338 \\
\textsc{arms\_number\_3} & 93.54\% & 93.47\% & -0.07\% & 1 \\
\textsc{arms\_number\_4} & 93.84\% & 85.45\% & -8.39\% & 1 \\
\textsc{arms\_number\_more\_than\_4} & 97.79\% & 97.51\% & -0.27\% & 1 \\
\textsc{arms\_number\_cant\_tell} & 86.13\% & 86.07\% & -0.06\% & 1 \\
\end{tabular}
    \caption{Quality metrics of morphological detailed features of galaxies from the inference dataset}
    \label{features}
\end{table*}

\section{Image-based catalogs of low-redshift SDSS galaxies 
classified \\ by five morphological classes and 34 morphological features}

Applying the CNN classifier to the inference galaxy dataset (low panel in Fig.~\ref{chart}), we took into account the following labels of galaxies: predictions of belonging to one of five morphological classes (Table~\ref{classes}) and 34 detailed morphological features (Table~\ref{features}). The augmentation procedures for image data of galaxies (decrease in stellar magnitude and correction of sizes) from the training dataset are described in our work [56].

\begin{figure}
	\includegraphics[width=1.0\textwidth]{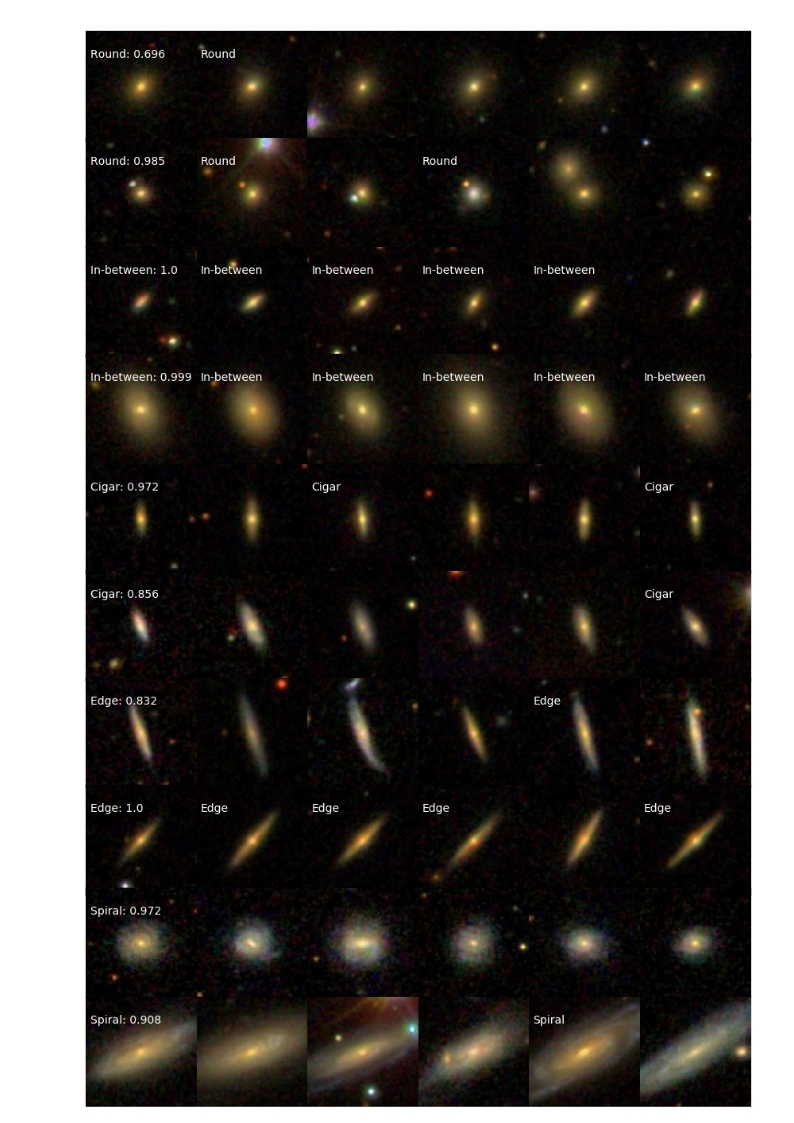}
    \caption{A set of inference galaxies (first column) with their five nearest neighbors from the GZ2 training dataset (the rest four columns). Each row represents the morphological class, which is intrinsic to the galaxy from the inference set. The number in the left upper corner of each image of the inference galaxies is a value of the corresponding probability of being this galaxy in a given class. Some classes of galaxies from the GZ2 training set are pointed out when possible (not all of the GZ2 galaxies fit our criteria for CNN classifier).}
    \label{composite}
\end{figure}

We have trained our CNN classifier and attained the overall accuracy of 89.3\,\% on the test set of $\sim 9000$ galaxies. It was obtained after splitting the training galaxy dataset (see, the distribution of predicted labels vs. true labels as the probabilities for the galaxy to belong to one of five morphological classes in Table~\ref{chart}). Assuming that a galaxy is in a certain morphological class if the probability is the highest one, we found that the inference dataset contains 27\,378 completely rounded, 59\,194 rounded in-between, 18\,862 cigar-shaped, 7\,831 edge-on, and 23\,119 spiral galaxies. 
So, a common classification of the studied sample of 315782 low-redshift SDSS galaxies with $m_{r}$ < 17.7 and $V_{LG}$>1500 km/s into five morphological classes consists of the following parts: 
\begin{itemize}
    \item 72\,738 galaxies from the training dataset and 143\,410 galaxies from the inference dataset, which have undergone the CNN model with the high adversarial score and the accuracy pointed in Table~\ref{chart}. It turned out 46\,846 completely rounded, 86\,515 rounded in-between, 22\,097 cigar-shaped, 13\,930 edge-on, and 41\,738 spiral galaxies.
    \item 105\,560 galaxies from the studied sample were not classified with the CNN model because of their low adversarial score (98\,534 galaxies) or technical reasons (7\,026 galaxies). We left the initial morphological classification for these galaxies into five classes as it was assigned in the GZ2 project.
\end{itemize}
Examples of the inference galaxies with their five nearest neighbors (in multi-label parametric space) from the GZ2 training dataset classified into five morphological classes with a given accuracy are shown in Fig.~\ref{composite}.

\begin{figure}
	\includegraphics[width=0.8\textwidth]{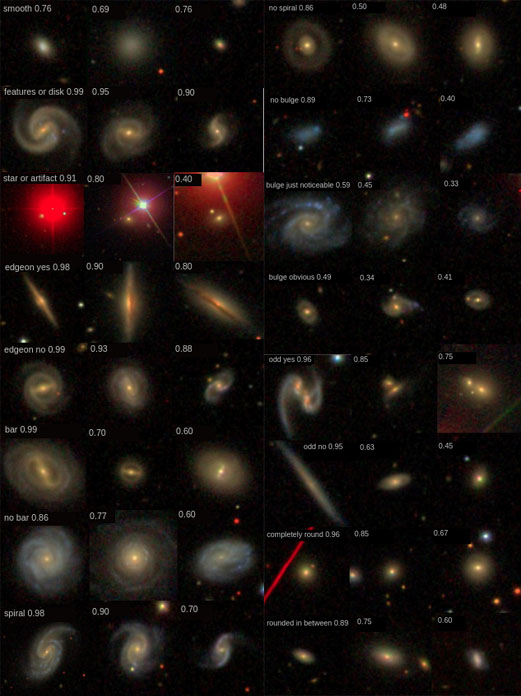}
	\centering
    \caption{The examples of galaxies with some morphological features (\textsc{smooth}, \textsc{features\_or\_disk}, \textsc{star\_or\_artifact}, \textsc{edgeon\_yes}, \textsc{edgeon\_no}, \textsc{bar}, \textsc{no\_bar}, \textsc{spiral}, \textsc{no\_spiral}, \textsc{no\_bulge}, \textsc{bulge\_just\_noticeable}, \textsc{bulge\_obvious}, \textsc{odd\_yes}, \textsc{odd\_no}, \textsc{completely\_round}, \textsc{rounded\_in\_between}, see, Table~\ref{features}) from the inference SDSS dataset with their two nearest neighbors from the GZ2 training dataset.}
    \label{arms}
\end{figure}

\begin{figure}
	\includegraphics[width=0.8\textwidth]{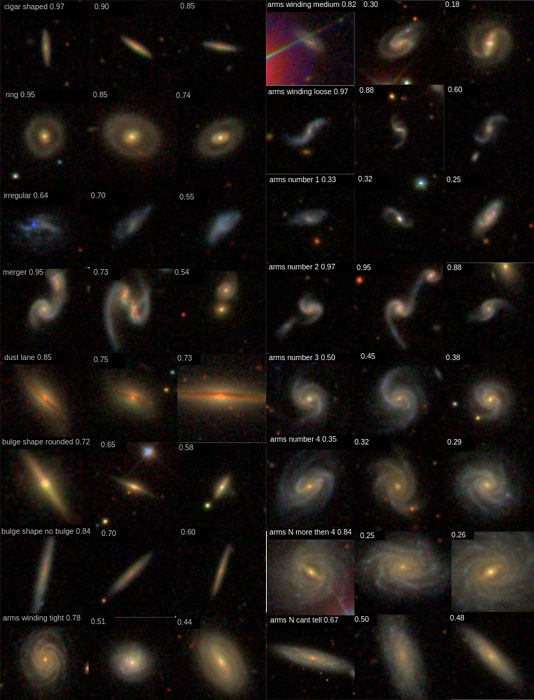}
	\centering
    \caption{The examples of galaxies with some morphological features (\textsc{cigar\_shaped}, \textsc{ring}, \textsc{irregular}, \textsc{merger}, \textsc{dust\_lane}, \textsc{bulge\_shape\_rounded}, \textsc{bulge\_shape\_no\_bulge}, \textsc{arms\_winding\_tight}, \textsc{arms\_winding\_medium}, \textsc{arms\_winding\_loose}, \textsc{arms\_number\_1}, \textsc{arms\_number\_2}, \textsc{arms\_number\_3}, \textsc{arms\_number\_4}, \textsc{arms\_number\_more\_than\_4}, \textsc{arms\_number\_cant\_tell}, see, Table~\ref{features}) from the inference SDSS dataset with their two nearest neighbors from the GZ2 training dataset.}
    \label{ring}
\end{figure}

Also, we determined the number of galaxies that passed the 0.5 cut-off for the acceptance of the detailed morphological features. The number of such galaxies with certain feature in the inference dataset is presented in Table~\ref{features}. The examples of inference galaxies with some of these morphological features (ring, bar, merger, irregular, etc.) with two nearest neighbors (in multi-label parametric space) from the GZ2 training dataset are shown in Fig.~\ref{arms} and Fig.~\ref{ring}. The mosaics of galaxies in these figures illustrate well how our CNN classifier work in finding the similar morphological features of various galaxies, for instance, to find the edge-on galaxies turned to the observer under the same angles (see seventh and eighth rows in Fig.~\ref{composite} or to find the galaxies with similar morphological features as a ring, bar or bulge (Fig.~\ref{ring}).

We used additional morphological parameters such as a bar or ring to predict the presence of these features in galaxies from our inference set. Because these features are not mutually exclusive, we introduce a different approach to determine their types. Namely, we assumed that the penultimate layer of a neural network consisting of 128 neurons must clearly characterize the galaxy. In its turn, the neighboring galaxies in the multidimensional parameter space must have the same characteristics. By determining the optimal number of nearest neighbors for the most accurate prediction and the optimal value of trimming the likelihood of signs, we test our hypothesis on a deferred GZ2 dataset and found a good confirmation.

\begin{figure}
	\includegraphics[width=1.0\textwidth]{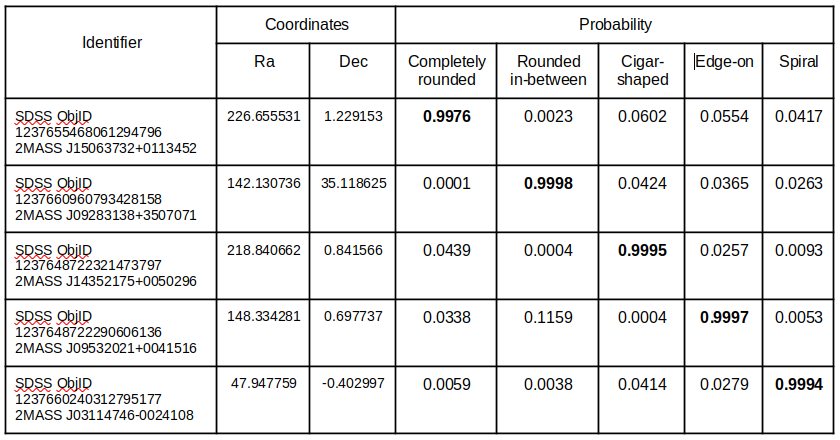}
    \caption{Examples of five galaxies from the Catalog of morphological classes of low-redshift galaxies from SDSS DR9, which have the highest probability to belong to the completely rounded, rounded in-between, cigar-shaped, edge-on, and spiral morphological classes.}
    \label{table3}
\end{figure}

\begin{figure}
	\includegraphics[width=1.0\textwidth]{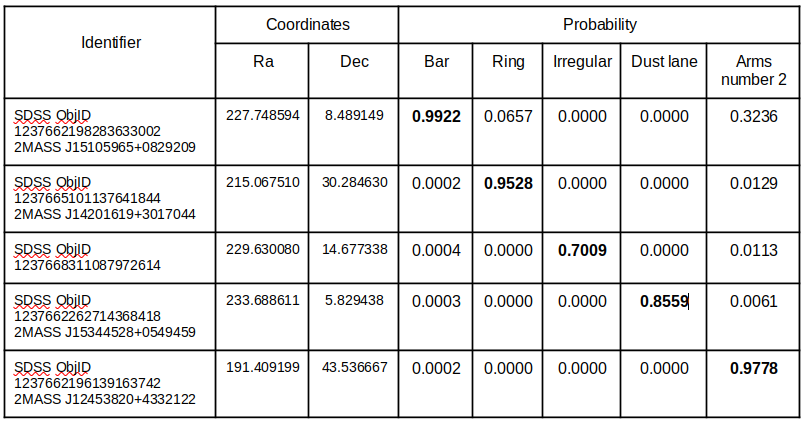}
    \caption{Examples of five galaxies from the Catalog of morphological classes of low-redshift galaxies from SDSS DR9, which have the highest probability to have ring, bar, irregular shape, dust lane, two spiral arms.}
    \label{table4}
\end{figure}

We created the catalog of 315,782 low-redshift galaxies from SDSS DR9, where morphological classes and detailed features were defined for the first time for 216148 galaxies by an image-based CNN classifier. For the rest of the galaxies (with the lower adversarial score), the initial morphological classification was re-assigned as in the GZ2 project. These new data will be added to our previous “Binary morphology SDSS galaxies catalog” [102] (see, also, SDSS DR9 Home Page \url{http://skyserver.sdss.org/dr9} and \url{https://cdsarc.cds.unistra.fr/viz-bin/cat/J/A+A/648/A122}). Examples of five galaxies from this catalogue, which have the highest probability to belong to the completely rounded, rounded in-between, cigar-shaped, edge-on, and spiral morphological classes, are listed in Fig.~\ref{table3}. Examples of five galaxies from this catalog, which have the highest probability to assign ring, bar, dust lane, and other morphological structural features, are listed in Fig.~\ref{table4}. 

\section{Discussion}
Classification of morphological types and features of galaxies is one of the cornerstones for extragalactic astronomy and observational cosmology. Galaxies of different morphological types are distributed non-uniformly across the sky and along the cosmological scale. The early-type galaxies predominate in the  central part of clusters. They also have larger masses, less gas, higher velocity dispersion, and diverse stellar population than the late-type galaxies (see, for example, \cite{Davies1983, Karachentseva1994, Cassata2011, Vulcani2011, Babyk2014, Tamburri2014, Dobrycheva2014, Dobrycheva2015,  Pulatova2015, Sybilska2017, Parikh2019, Puebla2020, Barchi2020, Reis2020, Vavilova2021c}. 

Astronomical surveys have accumulated a large number of galaxy images and data that need deep scientific exploration. For this purpose, it is very convenient to use relevant methods for a reliable automated galaxy morphological classification. There are many different options for sorting galaxies by type now, but each has its own drawbacks. For example, broad spectra of data are lost in spectroscopy classification because not all galaxies have spectra of good enough quality. Classifications based on the photometry give an error when trying to classify red spirals \cite{Vasylenko2019, Vavilova2020a}, i.e., galaxies with a high content of old stars or interacting galaxies which affect the photometric characteristics of each other \cite{Bottrell2019, Mezcua2014, Pearson2019}.
In favor of our choice of CNN as a basic model, we will mention several papers related to the image-based and photometry-based galaxy morphology classification with accentuating on the SDSS and Galaxy Zoo samples \cite{Lintott2008}.

\cite{Cabrera2018} explained how the human labeled biases in morphological photometry-based classification could be reduced through supervised ML. This coincides with our conclusion \cite{Vavilova2021a}, where we discuss which factors and properties of galaxies exactly affect the accuracy of supervised methods. In that paper, we concluded that one could not get the accuracy significantly exceeding 76\,\% when using the GZ2 data as a training set for ML with the photometry-based approach. One of the reasons is the attribution of irregular galaxies in the GZ2, which have the redder color indices, to the elliptical (early) type, and vice versa the elliptical galaxies with the bluer color indices to the spirals. In any case, the morphology obtained with the ML trained photometric parameters demonstrates significantly less bias than morphology based on citizen-science classifiers. This conclusion is also important for galaxies with a low surface brightness \cite{Du2019}. 

\cite{Cheng2020} used a set of ~2800 galaxies from Dark Energy Survey with visual classification from GZ1 to compare ML methods for galaxy classification: CNN, K-Nearest Neighbor, Logistic Regression, Support Vector Machine, Random Forest, and Neural Networks. They revealed that CNN was the most successful method in their study giving an accuracy of 99.4\,\% for the binary morphological classification of ellipticals and spirals. As for the combination of photometry and image galaxy SDSS data with Galaxy Zoo labels, we note the work by \cite{Hayat2020}, who applied self-supervised representation learning. \cite{Mittal2020} introduced the data augmentation-based MOrphological Classifier Galaxy (daMCOG CNN) using a convolutional neural network and obtained a testing accuracy of 98\,\%. Their datasets of 4614 images were collected from SDSS Image Gallery, Galaxy Zoo challenge, and Hubble Image Gallery.

\cite{Walmsley2020} used Bayesian CNNs and a novel generative model of Galaxy Zoo volunteer responses to infer posteriors for the visual morphology of galaxies. They show that training Bayesian CNNs with active learning requires up to 35–60\,\% fewer labeled galaxies depending on the morphological feature being classified. They concluded that in the synergy of human and machine intelligence, the Galaxy Zoo would be able to classify surveys of any conceivable scale, providing massive and detailed morphology catalogues to support research into galaxy evolution. These authors, in their next paper \cite{Walmsley2021}, used Galaxy Zoo data (SDSS DR8) and DECaLS data \cite{Dey2019} to provide the detailed visual morphology measurements in \textit{grb}-bands with Bayesian DL classifier for 314\,000 galaxies brighter than $m_{r}$=17.77 at $z <0.15$. Applying RGB image construction and various methods for the data processing, these authors were able to improve decision trees for the GZ2’ volunteer classification of galaxy morphological features.

Several useful catalogues were developed with the GZ classification scheme. \cite{Willett2013} issued a catalogue of morphological types from the GZ2 in the synergy with the SDSS DR7, which contains more than 16 million morphological classifications of 304\,122 galaxies and their finer morphological features (bulges, bars, and the shapes of edge-on disks as well as parameters of the relative strengths of galactic bulges and spiral arms). \cite{Simmons2017} cross-verified morphological features of $\sim$ 48\,000 galaxies from the CANDELS survey and the GZ project (clumpiness, bar instabilities, spiral structure, merging). It allowed them to create a list of galaxies with featureless disks at 1 $\leq z \leq$ 3, representing ``a dynamically warmer progenitor population to the settled disc galaxies seen at later epochs''. 

\cite{Dominguez2018} presented a morphological catalog for $\sim$ 670\,000 SDSS-galaxies in two options: T-type, related to the Hubble sequence, and GZ2 types. Their models with DL for the GZ2 type questions have the highest accuracy (>97\,\%) when applied to a test dataset with the same characteristics as the one used for training dataset. In the recent work \cite{Dominguez2018}, they presented the MaNGA Deep Learning Morphological Value Added Catalogue as a part of the SDSS DR17. This catalog includes a number of morphological properties: e.g. a T-Type, a finer separation between elliptical and S0, as well as the identification of edge-on/barred galaxies and a separation between early/late types.

\cite{Vega2021} presented morphological classifications of $\sim$ 27 million galaxies from the Dark Energy Survey (DES DR1). They provided a CNN model to classify these galaxies by early and late types (accuracy $\sim$ 87\,\%) as well as by face-on and edge-on galaxies (accuracy 73\,\%). \cite{Dominguez2021} in their work related to the algorithm for preparing this largest automated morphological catalogue up to date,  describe how their trained SDSS image data of galaxies were transferred on the DES images. They also modeled fainter objects by simulating what the brighter objects with well-determined classifications would look like if they were at higher redshifts. This is the same as we applied in our approach.

The results mentioned above are quite comparable in accuracy with each other in determining the morphological peculiarities of galaxies and are in good agreement with our results. It's evident that CNN models are effective enough for the image-based classification of galaxy morphological features. 

In general, our method shows a satisfactory level of the morphological classification performance, attaining more than 90\,\% of accuracy for most morphological classes (Table~\ref{classes}). Such value of the accuracy is in good agreement with the accuracy obtained in the work by \cite{Walmsley2020}, who used Bayesian CNN to study Galaxy Zoo volunteer responses and achieved coverage errors of 11.8\,\% within a vote fraction deviation of 0.2. As well as with work by \cite{Gauthier2016}, who applied both supervised and unsupervised methods to study the Galaxy Zoo dataset of 61578 pre-classified galaxies (spiral, elliptical, round, disk). They found that the variation of galaxy images is correlated with brightness and eccentricity, and the accuracy for galaxies to be associated with each of these four classes is about 94\,\%. 

As for the classification of galaxies by their detailed structural morphological features, our CNN model gives the accuracy in the range of 92–99\,\% depending on features, a number of galaxies with the given feature in the inference dataset, and, of course, the galaxy image quality (Table~\ref{features}). As a result, for the first time, we assigned 34 morphological detailed features for more than 160\,000 low-redshift galaxies with $m_{r}$ <17.7 from the SDSS DR9, which had the highest adversarial score by our CNN classifier. If we compare our result with the largest galaxy morphological catalogue presented by \cite{Vega2021}, where the face-on and edge-on galaxies were classified by their images with an accuracy of 73\,\%, we constitute that our CNN model gives a more significant output.

\section{Conclusion}
We developed a CNN model for image-based galaxy morphological classification. The studied sample of 315\,782 galaxies with $m_{r}$<17.7 from SDSS DR9 at 0.02<z<0.1 is overlapped with the data from the Galaxy Zoo 2 (GZ2). It allowed us to divide it into two datasets: ``inference dataset'' of 143,410 galaxies, which do not match with the GZ2 dataset; ``training dataset'' of 172\,372 galaxies, which match with galaxies from our studied sample.

To develop the CNN model based on the images of galaxies, we used the GZ2 assigned labels for five morphological classes by shape and for 34 detailed structural morphological features of galaxies. We revealed that adversarial validation is very helpful when the labeled datasets are biased in magnitude distribution for the training dataset, and such a difference could bias the final prediction of the classifier on the inference dataset. To avoid this problem, we applied the adversarial validation method for analyzing the homogeneity of these two datasets and for modeling fainter galaxies. As a result, the galaxies were selected from the training dataset with the highest adversarial score that are most closely coincided with the inference dataset, and the images were normalized to be similar. Our CNN classifier has demonstrated >93\,\% of accuracy for rounded in-between, edge-on, and spiral morphological classes, 83\,\% for the completely rounded, and 75\,\% for the cigar-shaped galaxies. Assuming that a galaxy is in a certain morphological class if the probability is the highest one, we found that the inference dataset contains 27\,378 completely rounded, 59\,194 rounded in-between, 18,862 cigar-shaped, 7\,831 edge-on, and 23\,119 spiral galaxies.

As for the detailed structural features of galaxies, we worked with the inference dataset of 160\,471 galaxies and with 34 galaxy morphological features (bar, ring, various number of spiral arms, disks, dust lane, merger etc.) as they are labeled by the GZ2’s volunteers. We used ROC AUC (Area Under Receiver Operating Characteristic Curve) as the binary classification quality metric, which gives accuracy scores for the model trained with adversarial augmentations and for the model trained without these augmentations. Our CNN model provides the accuracy in the range of 92–99\,\% depending on the features, the number of galaxies with the given feature in the inference dataset, and, of course, the galaxy image quality. To ensure robustness and reliability, we have also visually inspected the galaxy images. As a result, for the first time, we assigned 34 morphological detailed features for more than 160\,000 low-redshift galaxies with $m_{r}$ <17.7 from the SDSS DR9, which have the highest adversarial score by our CNN classifier.

In general, we created the catalogue of 315782 low-redshift galaxies from SDSS DR9, where morphological classes and detailed features were defined, for the first time, for 216\,148 galaxies by the image-based CNN classifier. For the rest of galaxies (with the lower adversarial score), the initial morphological classification was re-assigned as in the GZ2 project. These catalogues can be accessed through the VizieR CDS platform. A vector representation of the probability distribution of a galaxy having one or another feature (the penultimate layer of our CNN model) can be founded at the Ukrainian Virtual Observatory \cite{Vavilova2012} web-site (\url{http://ukr-vo.org/catalogues}). This will be of interest to those who will study the similarities between galaxies in more detail. Our approach to the image data augmentation can be applied as the mathematical tools in tasks of positional and photometrical processing CCD frames, archive astroplates in various bands, transient objects, artifacts etc. \cite{Djorgovski2016, Morello2018, Ruhe2020, Savanevych2018, Villarroel2020}. The proposed CNN model allows solving a bunch of galaxy classification problems, for example, such as a quick selection of galaxies with a bar, bulge, or ring for their subsequent analysis. Our approach consumes the time at the stage of preliminary preparation of the studied galaxy dataset and can be useful for further studies of the morphology, image, photometry, and spectroscopic data of galaxies.

\begin{acknowledgements}
We are grateful to the referee for useful comments that allowed us to present the results of our study more fully. The use of the SDSS \citep{Ahn2012, Blanton2017} and SAO/NASA Astrophysics Data System was extensively applicable. The authors thank the Galaxy Zoo team.

This work was done in the frame of the budgetary program «Support for the development of priority fields of scientific research» of the NAS of Ukraine (CPCEL 6541230). 
\end{acknowledgements}

\bibliographystyle{aa} 
\bibliography{vavilova-et-al-2022} 

\end{document}